\documentclass[12pt]{article}
\usepackage{epsfig,amssymb,euscript,bbold}
\usepackage{amsfonts}
\usepackage{eufrak}
\usepackage{mathrsfs}
\usepackage{graphicx}
\usepackage{epsfig}
\usepackage{bm}
\usepackage{amsmath}
\def\be#1\ee{\begin{equation}#1\end{equation}}
\newcommand{\bea}{\begin{eqnarray}}
\newcommand{\eea}{\end{eqnarray}}
\newcommand{\ba}{\begin{array}}

\newcommand{\ea}{\end{array}}

\def\bbox{{\,\lower0.9pt\vbox{\hrule \hbox{\vrule height 0.2 cm
\hskip 0.2 cm \vrule height 0.2 cm}\hrule}\,}}
\newcommand{\dsl}{\pa \kern-0.5em /}

\newcommand{\nn}{\nonumber \\}

\def\w{\wedge}
\def\mbb{\mathbb{R}}

\def\l{\lambda}

\def\r{\rho}
\def\a{\alpha}

\def\ds{\raise.15ex\hbox{/}\kern-.57em\partial}
\def\Ds{\,\raise.15ex\hbox{/}\mkern-13.5mu D}
\newcommand{\dd}{\mathrm{d}}

\newcommand{\ee}{\mathrm{e}}

\begin{document}

\makeatletter
\renewcommand{\theequation}{\thesection.\arabic{equation}}
\@addtoreset{equation}{section}
\makeatother

\baselineskip 18pt

\begin{titlepage}

\vfill

\begin{flushright}
Imperial/TP/2007/OC/02\\
\end{flushright}

\vfill

\begin{center}
   \baselineskip=16pt
   {\Large\bf Inverting geometric transitions: explicit Calabi-Yau
     metrics for the Maldacena-Nu\~{n}ez solutions}
   \vskip 2cm
      Ois\'{\i}n A. P. Mac Conamhna 
   \vskip .6cm
      \begin{small}
      \textit{Blackett Laboratory, Imperial College\\
        London, SW7 2AZ, U.K;}
\vskip 0.5cm
\textit{The Institute for Mathematical
        Sciences,\\Imperial College, London SW7 2PG, UK.}
        \end{small}
   \end{center}

 \vskip .15 in
\begin{center}
{\texttt{o.macconamhna@imperial.ac.uk}}
\end{center}

\begin{center}
\textbf{Abstract}
\end{center}

\begin{quote}
Explicit Calabi-Yau metrics are derived that are argued to map to the
Maldacena-Nu\~{n}ez $AdS$ solutions of M-theory and IIB under 
geometric transitions. In 
each case the metrics are singular where a $H^2$ K\"{a}hler two-cycle
degenerates but are otherwise smooth. They are derived as the most
general Calabi-Yau solutions of an ansatz for the supergravity
description of branes wrapped on K\"{a}hler two-cycles. The ansatz is
inspired by re-writing the $AdS$ solutions, and the structure defined
by half their Killing spinors, in this form. The
world-volume theories of fractional branes wrapped at the
singularities of these metrics are proposed as the duals of the $AdS$
solutions. The existence of supergravity solutions interpolating
between the $AdS$ and Calabi-Yau metrics is conjectured and their
boundary conditions briefly discussed. 

\end{quote}

\vfill

\end{titlepage}

\setcounter{equation}{0}

\section{Introduction and main idea}
The AdS/CFT correspondence \cite{malda} is best understood for D3
branes at the apex of a Calabi-Yau cone. There are two ways in which
we know how to
think about this system. One is in terms of open string theory and
probe $D3$ branes on
the singular Calabi-Yau; at low energies, one gets a
four-dimensional conformal field theory, at weak 't Hooft coupling, on
the brane worldvolume. The other is in terms of closed string theory
on the product of $AdS_5$ with a Sasaki-Einstein manifold; by the
AdS/CFT correspondence, this is the same as the CFT at strong 't Hooft
coupling. The classical link between the two geometries is a smooth
supergravity solution, preserving half their supersymmetries, that
interpolates between them; the Calabi-Yau singularity is excised and
replaced with an $AdS$ horizon at infinite proper distance. In this
sense the branes are said to induce a geometric transition: they resolve
(rather, remove to infinity) the singularity of the Calabi-Yau manifold. The
geometrical data of both the Calabi-Yau and the Sasaki-Einstein
manifold are encoded in the CFT (at weak and strong coupling,
respectively), so interpolating the 't Hooft coupling in the CFT gives a quantum
definition of the geometric transition. The dictionary - encoding and
decoding Calabi-Yau and Sasaki-Einstein data in the CFT at weak and
strong coupling, respectively - has been worked out in detail in
beautiful work for $\mbb^6$, the conifold, and the $Y^{p,q}$ metrics
\cite{kleb}-\cite{6}. 

Since the work of Maldacena and Nu\~{n}ez \cite{mn}, we know that
there are many other ways in which Anti-de Sitter geometries can be
related to special holonomy manifolds and conformal quantum
theories. In \cite{mn}, three 
$AdS$ solutions of M- and string theory were constructed: two $AdS_5$
solutions in eleven dimensions, with respectively sixteen and eight
Killing spinors, and an $AdS_3$ solution admitting eight Killing
spinors in IIB\footnote{These solutions will be denoted by
  MN(I), MN(II) and MN(III) respectively.}. These were interpreted as
arising, in the near-horizon limit, from branes wrapping $H^2$ K\"{a}hler
two-cycles in, respectively, Calabi-Yau two-, three-, and
three-folds. The dual conformal field theories are $\mathcal{N}=2$ and
$\mathcal{N}=1$ in four dimensions, and $N=(2,2)$ in three
dimensions. Since this work, it has been found that there exist $AdS$
solutions associated to all types of calibrated cycles in all types of
special holonomy manifold of dimension ten or less; for example,
\cite{acharya}-\cite{gcal}. The CFTs dual to $AdS$ manifolds of this
type define quantum gravity theories for calibrated geometries. In line
with the intuition gained from branes at conical singularities, one
would expect that the CFTs could be realised, at weak coupling, as the
world-volume theories of fractional probe branes, wrapped on
degenerating calibrated cycles in singular special holonomy
manifolds. Such a system is likewise expected to undego a geometric
transition, with the singularity excised and replaced with an $AdS$
region. Classically, there should be a supergravity solution
interpolating between the Calabi-Yau and $AdS$ geometries.

Our understanding of AdS/CFT for wrapped branes is much more
rudimentary than for branes at conical singularities. Chief among the
obstacles has been the inability to move beyond the near-horizon limit;
typically, only the $AdS$ geometries are known. The lesson from branes
on cones is that in order to get a real handle on field theory {\it
  dynamics} - to write down the particle content and superpotential
for a dual of a specific $AdS$ solution - the associated Calabi-Yau
geometry must be known. The main point of this paper is to give a way
of associating a special holonomy metric to an $AdS$ metric,
illustrated for the Maldacena-Nunez solutions. The main assumption of
this paper is the existence of a supersymmetric supergravity solution
interpolating 
between a special holonomy manifold and an $AdS$ spacetime when there
exists an AdS/CFT dual. Roughly, an interpolating solution should be a
metric and a flux admitting two distinct limits in which the supersymmetry 
doubles, with the metric becoming Calabi-Yau in one limit and $AdS$ in
the other. More formally, we can think of the metric and flux of an
interpolating solution as providing a smooth and smoothly-invertible map $f$
\bea
f:\mbox{Special Holonomy}\rightarrow AdS.
\eea
We take this as a definition of what is meant in this paper by a geometric
transition. It is a purely classical definition; in more
physical terms, such a map gives the full supergravity desription of a
wrapped brane. But if a CFT dual can be identified, the map can be
promoted to the quantum level; the CFT itself provides the map, with
the 't Hooft coupling the interpolating parameter.

The equations that interpolating solutions should satisfy are known,
through various symmetry arguments. An important property of these
solutions is that they should admit a global reduction of their frame
bundle, to a sub-bundle of the appropriate structure \cite{wrap},
\cite{pau}. For example, in the supergravity description of M5-branes
wrapped on K\"{a}hler two-cycles in Calabi-Yau two-folds - maps
$f:CY_2\rightarrow AdS_5$ - the global structure of an interpolating
solution is $SU(2)$. The structure is defined by eight Killing spinors,
or alternatively, an almost complex structure $J$ and a $(2,0)$ form
$\Omega$. The truncation of eleven-dimensional supergravity to this
frame bundle was first worked out by Fayyazuddin and Smith
\cite{fz} (see also \cite{fz1}, \cite{wrap}). The metric and flux are 
\bea
\dd s^2&=&L^{-1}\dd s^2(\mbb^{1,3})+\dd s^2(\mathcal{M}_4)+L^2[\dd
t^2+t^2\dd s^2(S^2)],\nn
\star_7F&=&L^2\dd(L^{-2}J).
\eea
Here and throughout we follow all conventions and orientations of
\cite{wrap}.  The Minkowski isometries are isometries of the full solution, and
$\mathcal{M}_4$ admits a globally-defined $SU(2)$ structure. The
structure is constrained by the Fayyazuddin-Smith equations:
\bea
\dd(L^{-1/2}\Omega)&=&0,\nn\label{torsion}
\dd t\w\mbox{Vol}[S^2]\w\dd(LJ)&=&0.
\eea
Eleven-dimensional supergravity, in this truncation, reduces to the
torsion conditions \eqref{torsion} and the four-form Bianchi identity.  

To the knowledge of the author, no interpolating solutions of these
equations, or their analogues in other contexts, are known. However in
recent work \cite{wrap}, \cite{eoin}, \cite{pau}, it has been shown
how the supersymmetry conditions for general classes of
supersymmetric $AdS$ solutions of M-theory (including all known
examples) can be derived from such equations. In particular, in
\cite{wrap} it was shown that the conditions of Lin, Lunin and
Maldacena \cite{llm} for half-BPS $AdS_5$ solutions can be derived from the
Fayyazuddin-Smith equations. It follows that any solution of the
LLM conditions can be re-written as a solution of the Fayyazudin-Smith
equations; and similarly for every other $AdS$ solution covered by
\cite{wrap}, \cite{eoin}, \cite{pau}. Applying this procedure to the
MN(I) solution, 
we will see in the next section that it may be re-written in the form
\bea
\dd s^2&=&L^{-1}\Big[\dd s^2(\mbb^{1,3})+\frac{F}{2}\dd
s^2(H^2)\Big]+L^2\Big[F^{-1}\Big(\dd u^2+u^2(\dd\psi-P)^2\Big)+\dd t^2+t^2\dd
s^2(S^2)\Big],\nn\dd P&=&\mbox{Vol}[H^2],
\eea
for particular determined functions $F(u,t),L(u,t)$ and a particular
choice of frame which will be discussed in detail. We use this form of
the $AdS$ solution as a guide to what the inverse geometric transition
$f^{-1}:\mbox{MN(I)}\rightarrow CY_2$ should be. Clearly, it should
respect the topological structure of MN(I); the simplest choice, which
we make, is that $f$ is given by a solution $F(u,t),L(u,t)$, of the
Fayyazuddin-Smith equations. With this metric and the frame of section
2, they reduce to
\bea
\frac{1}{t^2}\partial_t\Big(t^2\partial_tF\Big)&=&-u\partial_u\Big(\frac{F}{u}\partial_uF\Big),\nn\label{geo}
L^3&=&-\frac{1}{4u}\partial_u(F^2).
\eea
An interpolating solution of these equations has not been
found. However, assuming one exists, the general Calabi-Yau solution
of \eqref{geo} is the image of MN(I) under $f^{-1}$. Up to an overall
scale, the general Calabi-Yau solution is $L=1$ and
\bea
\dd s_4^2=\frac{\dd
  R^2}{\left(\frac{1}{R^4}-1\right)}+\frac{R^2}{4}\left[\dd
  s^2(H^2)+\left(\frac{1}{R^4}-1\right)(\dd \psi-P)^2\right].
\eea
The range of $R$ is $[-1,0)$ or $(0,1]$. As expected, the metric is
singular, where the K\"{a}hler two-cycle $H^2$ degenerates. The
singularity, at $R=0$, is at finite
proper distance. The metric is non-singular at the $H^2$ bolt as
$R^4\rightarrow 1$, if $\psi$ has period $2\pi$; we will see in the
next section that this is precisely the periodicity
that is inherited through $f^{-1}$ from MN(I). Some additional
evidence that this Calabi-Yau is a sensible candidate comes from the
following. Every $AdS_5$ solution of the LLM conditions, including
MN(I), is completely determined by a solution of the three-dimensional
continuous Toda equation. There also exists a class of Calabi-Yau
two-folds that is completely determined by a solution of the
three-dimensional continuous Toda equation. This is such a Calabi-Yau
metric, and furthermore it is given by the same solution of the Toda
equation as MN(I). Toda-Calabi-Yau metrics have been obtained in this
context before as scaling limits of the 1/2-BPS $AdS_5$ metrics
\cite{llm}, \cite{lunin}. Here this metric is obtained in a different
way, as a solution of the 1/4-BPS Fayyazuddin-Smith equations. It will
be interesting to see how these procedures are related.

The world-volume theory of fractional M5-branes wrapped at the
singularity of this metric (whatever it might be) is proposed as the
quantum dual of MN(I). Though
the geometry is non-compact, this is not necessarily problematic, as
the field theory should only encode
oscillations in the directions transverse to the brane, purely in the
fibre; and the fibre has finite proper volume. The cycle may in any
event be rendered compact by taking a freely-acting quotient by a
discrete subgroup of its isometry group. The Calabi-Yau will
still be noncompact, because of the singularity. 

In a similar vein, we obtain candidate Calabi-Yaus for inverse
geometric transitions from MN(II) and MN(III). For MN(II), to be
discussed in detail in section 3, the first
step is to use the results of \cite{ypq}, \cite{wrap} to write it in
the form
\bea
\dd s^2&=&L^{-1}\left[\dd s^2(\mbb^{1,3})+\frac{F_1F_2}{3}\dd
  s^2(H^2)\right]\nn&&+L^2\left[F_1^{-1}\left(\dd
    u^2+\frac{u^2}{4}(\dd\psi+P-P')^2\right)+ F_2^{-1}\frac{u^2}{4}\dd
  s^2(S^2)+\dd t^2\right].
\eea
with $\dd P=\mbox{Vol}[S^2]$ and $\dd P'=\mbox{Vol}[H^2]$. Then, 
letting $L,F_1,F_2$ be arbitary functions of $u,t$, the general
Calabi-Yau three-fold solution\footnote{With suitable regularity
  properties, to be discussed in detail in section 3.} is, up to an
overall scale,
\bea\label{cyy}
\dd s^2&=&\frac{1}{2}(1+\sin\xi)\dd
s^2(H^2)+\frac{\cos^2\xi}{2(1+\sin\xi)}\dd
s^2(S^2)+\frac{1}{\cos^2\xi}\Big(\dd
R^2+R^2(\dd\psi+P-P')^2\Big), 
\eea
where $\sin\xi$ is a root of the cubic equation
\bea
-\frac{1}{3}\sin^3\xi+\sin\xi=\frac{2}{3}-R^2.
\eea
The metric is singular, as expected, at $\xi=-\pi/2$, $R=2/\sqrt{3}$,
where the $H^2$ cycle degenerates. The metric is smooth at
$\xi=\pi/2$, which coincides with $R=0$; there an $S^3$ smoothly
degenerates, provided that $\psi$ has the $4\pi$ periodicity it
inherits from MN(II) under $f^{-1}$. The quantum dual of MN(II) is
proposed to be the worldvolume theory of fractional M5s wrapped at the
singularity of this metric. 

The MN(III) solution comes from D3 branes wrapped
on a K\"{a}hler $H^2$ cycle in a Calabi-Yau three-fold; the geometric
transition is $CY_3\rightarrow AdS_3$ in IIB. The discussion in this
case proceeds along very much the same lines as for MN(II), and will
be reported in detail elsewhere \cite{us}. The image of MN(III) under
the inverse transition is again the Calabi-Yau \eqref{cyy}. The dual
field theory is proposed to be the world-volume theory of D3 branes
wrapped at the singularity. It seems
that both M5 branes and D3 branes can probe the singularity of this
manifold; the quantum descriptions are respectively four- and
two-dimensional conformal theories. In the IIB description, given the
metric, it might be possible to construct the field theory with
existing techniques.

The remainder of this paper is organised as follows. Section 2 is
devoted to MN(I) and its Calabi-Yau image. Section 3 repeats the
analysis of section 2 for MN(II). Section four contains conclusions
and outlook, and also some discussion
of the boundary conditions for interpolating solutions for the
MN/Calabi-Yau pairs.

\section{The $\mathcal{N}=2$ M-theory solution}
To begin, we will review the $\mathcal{N}=2$ $AdS_5$ geometry of \cite{mn}
in some detail. The metric is given by
\bea\label{mtt}
\dd s^2&=&\frac{1}{\l}\left[\dd s^2(AdS_5)+\frac{1}{2}\dd
  s^2(H^2)+(1-\l^3\r^2)(\dd
  \psi-P)^2+\frac{\l^3}{4}\left(\frac{\dd\r^2}{1-\l^3\r^2}+\r^2\dd
    s^2(S^2)\right)\right],\nn&&
\eea 
where
\bea
\l^3&=&\frac{8}{1+4\r^2},\nn
\dd P&=&\mbox{Vol}[H^2].
\eea
Here and throughout we denote by $\dd s^2(\mathcal{M})$ the metric of
unit radius of curvature on $\mathcal{M}$. The range of the coordinate
$\rho$ is either $\rho\in[-1/2,0]$ or $\rho\in[0,1/2]$. At $\rho=0$,
in either branch, the R-symmetry $S^2$ smoothly
degenerates\footnote{The R-symmetry of the dual theory is $SU(2)\times
  U(1)$.}. As
$\rho\rightarrow\pm1/2$, the R-symmetry $U(1)$, with coordinate
$\psi$, smoothly degenerates, provided that $\psi$ is identified with
period $2\pi$. Henceforth we will take $\rho$ to be non-negative. 

This manifold, as a solution of eleven-dimensional supergravity,
admits sixteen Killing spinors. The Killing spinors may be used to
define an identity 
structure - a preferred frame associated to them. The structure is
discussed in detail in \cite{wrap}. We choose
coordinates for the preferred frame according to
\bea
e^1+ie^2&=&\frac{1}{\sqrt{2\l}}e^{i\psi}(\dd\mu+i\sinh\mu\dd\beta),\nn
e^3&=&\sqrt{\frac{1-\l^3\r^2}{\l}}(\dd\psi-\cosh\mu\dd\beta),\nn
\hat{\rho}&=&\frac{\l\dd\rho}{2\sqrt{1-\l^3\r^2}},\nn
\hat{r}&=&\l^{-1/2}\dd r,
\eea
where we have chosen Poincar\'{e} coordinates on $AdS$,
\bea
\dd s^2(AdS_5)=e^{-2r}\dd s^2(\mbb^{1,3})+\dd r^2.
\eea
The remaining directions play no r$\hat{\mbox{o}}$le in the rest of the
discussion. 

The MN(I) solution is a particular case of a broader class of half-BPS
$AdS_5$ solutions which are completely determined by a solution of the
three-dimensional continuous Toda equation\footnote{It is strongly believed, at least by the
  author, that all half-BPS $AdS_5$ solutions of M-theory are of this
  form.}. The Toda equation,
\bea\label{tod}
\nabla^2_{\mbb^2}D+\partial_{\rho}^2e^D=0,
\eea
may be viewed as a three-dimensional Laplace equation, 
\bea
\nabla_3^2D=0,
\eea
on a three-manifold with metric
\bea
\dd s^2=\dd\rho^2+e^D\dd s^2(\mbb^2).
\eea
The metric on every half-BPS $AdS_5$ solution of eleven dimensional
supergravity determined by the Toda equation may be written as follows
\cite{llm}: 
\bea\label{MN(I)}
\dd s^2&=&\frac{1}{\l}\Big[\dd
  s^2(AdS_5)+(1-\l^3\r^2)(\dd\psi+V)^2+\frac{\l^3}{4}\Big(\frac{1}{1-\l^3\r^2}[\dd\r^2+e^D\dd
      s^2(\mbb^2)]+\r^2\dd s^2(S^2)\Big)\Big],\nn
\eea
where
\bea
\l^3&=&\frac{-\partial_{\r}D}{\rho(1-\rho\partial_{\rho}D)},\\
V&=&\frac{1}{2}\star_2\dd_2D,
\eea
where $\dd_2$ is the exterior derivative restricted to $\mbb^2$, and
$D$ solves \eqref{tod}. The MN(I) solution is given by
\bea
e^D=\frac{1}{4x_1^2}(1-4\rho^2).
\eea

In order to make the relationship between the MN(I) geometry and
wrapped branes more concrete, we now want to exhibit it as a solution
of the Fayazzuddin-Smith equations. The essential point is that in
addition to its identity structure, MN(I) also admits an $SU(2)$
structure, defined by {\it half} its Killing spinors, which indeed
solves the Fayyazuddin-Smith equations. In \cite{wrap} it was shown how to
obtain an arbitrary Toda-$AdS_5$ manifold as a 1/4 BPS solution of the
wrapped brane conditions by constructing its $SU(2)$ structure. Here
we will apply these general results to the specific case of interest.

The canonical frame of the identity structure is related to
the canonical frame of the $SU(2)$ structure by a local rotation. If
we define the Minkowski frame $e^a,e^4,\hat{t}=L\dd t$, $a=1,2,3$,
with $e^a,e^4$ a basis for $\mathcal{M}_4$, the relationship between the
``$AdS$'' and ``Minkowski'' frames is given by
\bea
e^a_{Mink}&=&e^a_{AdS},\nn
e^4&=&\cos\theta\hat{\rho}+\sin\theta\hat{r},\nn\label{kf}
\hat{t}&=&-\sin\theta\hat{\r}+\cos\theta\hat{r}.
\eea
For more details of this procedure, which seems to be a universally
applicable way of writing $AdS$ manifolds in a wrapped brane form, the
reader is referred to \cite{wrap}, \cite{eoin}, \cite{pau}\footnote{The frame
rotation, as a way of deriving warped $AdS_{d+2}$ supersymmetry conditions
from warped $\mbb^{1,d}$ supersymmetry conditions, was first employed
in \cite{ypq}.}. In the
case at hand, the rotation angle is related to the $AdS$ warp factor
and the coordinate $\rho$ by 
\bea
\cos\theta=\l^{3/2}\r,
\eea
and also the warp factors are related by
\bea
L=\l e^{2r}.
\eea     
Near $\rho=0$, the $AdS$ radial direction aligns with $\pm e^4$. Near
$\rho=1/2$, it aligns with $\hat{t}$. Since we know everything on the
right-hand side of \eqref{kf}, we can 
see that
\bea
e^4&=&Le^{-r}\dd\left(-\sqrt{\frac{1-4\r^2}{8}}e^{-r}\right),\nn
\hat{t}&=&L\dd\left(-\frac{\rho}{2}e^{-2r}\right).
\eea
Therefore defining the Minkowski-frame coordinates
\bea
u&=&-\sqrt{\frac{1-4\r^2}{8}}e^{-r},\nn
t&=&-\frac{\rho}{2}e^{-2r},
\eea 
we can re-write the $AdS_5$ solution as
\bea
\dd s^2&=&L^{-1}\Big[\dd s^2(\mbb^{1,3})+\frac{F}{2}\dd
s^2(H^2)\Big]+L^2\Big[F^{-1}\Big(\dd u^2+u^2(\dd\psi-P)^2\Big)+\dd t^2+t^2\dd
s^2(S^2)\Big],\nn&&
\eea
where $F=e^{2r}$ is determined by a root of the quadratic
\bea\label{quad}
2t^2e^{4r}+u^2e^{2r}-\frac{1}{8}=0.
\eea
One of these roots is always negative, so we choose the other, which
is always positive:
\bea
F=\frac{u^2}{4t^2}\left(-1+\sqrt{1+t^2/u^4}\right).
\eea
The warp-factor in the Minkowski frame is 
\bea
L^3&=&\frac{u^2}{\sqrt{1+t^2/u^4}}\left(\frac{-1+\sqrt{1+t^2/u^4}}{4t^2}\right)^2.
\eea
The canonical frame for the $SU(2)$ structure, re-written in terms of
the new coordinates, is
\bea
e^1+ie^2&=&\sqrt{\frac{F}{2L}}e^{i\psi}(\dd\mu+i\sinh\mu\dd\beta),\nn
e^3&=&-\frac{Lu}{\sqrt{F}}(\dd\psi-\cosh\mu\dd\beta),\nn
e^4&=&\frac{L}{\sqrt{F}}\dd u,\nn\label{fframe}
\hat{t}&=&L\dd t,
\eea
with a minus sign in the second equation because of the definition of
$u$. The $SU(2)$ structure then takes the standard form:
\bea
J&=&e^{12}+e^{34},\nn
\Omega&=&(e^1+ie^2)(e^3+ie^4),
\eea
and it may now be verified by explicit computation that it satisfies
the Fayyazudin-Smith equations. This was, of course, guaranteed by the
construction, but it serves as a consistency check. Having obtained the
$SU(2)$ structure of MN(I), it is now an
obvious thing to use it as an ansatz for further, topologically related,
solutions of the Fayyazuddin-Smith equations. To this end, we let
$F$ and $L$ be arbitrary functions of $u,t$, and insert the frame
\eqref{fframe} into the $SU(2)$ torsion conditions and Bianchi identity. They
reduce to the single non-linear second order pde for F :
\bea
\frac{1}{t^2}\partial_t\Big(t^2\partial_tF\Big)=-u\partial_u\Big(\frac{F}{u}\partial_uF\Big).
\eea
Given a
solution of this equation, $L$ is then determined by 
\bea
L^3=-\frac{1}{4u}\partial_u(F^2).
\eea
As a purely mathematical aside, we observe that the other root of the
quadratic \eqref{quad} is also a solution of these equations. But of
particular interest is the most general Calabi-Yau solution of this 
system. It may be most easily determined by imposing
$L=\mbox{constant}$ and closure of $J$. The $(2,0)$ form $L^{-1/2}\Omega$ is
always closed with this ansatz. The general Calabi-Yau solution is
\bea
F^2=a+bu^2.
\eea
For a metric of the right signature, we must have $a>0$, $b<0$. By rescaling,
we can set $b=-2$, so that $L=1$ (up to an overall scale in the
eleven-dimensional metric). This Calabi-Yau is diffeomorphic to a Toda-Calabi-Yau, as may
be seen by performing the coordinate transformation
\bea
16U^2=a-2u^2.
\eea
Defining $A^2=a/4$, the metric becomes
\bea
\dd s^2=\frac{4}{\partial_uD}(\dd \a+V)^2+\partial_uD(\dd u^2+e^D\dd s^2(\mathbb{R}^2)),
\eea
where 
\bea
e^D=\frac{1}{4x_1^2}(A^2-4U^2),
\eea
which, modulo the constant, is the same solution of the Toda equation
as that determining 
MN(I). An alternative form of the metric, reminiscent of
Eguchi-Hanson, is given by choosing the coordinate
\bea
R^2=\frac{1}{a^{1/4}}\sqrt{2a-4u^2}.
\eea
Up to an overall scale the metric becomes
\bea\label{eguchi}
\dd s^2=\frac{\dd
  R^2}{\left(\frac{1}{R^4}-1\right)}+\frac{R^2}{4}\left[\dd
  s^2(H^2)+\left(\frac{1}{R^4}-1\right)(\dd \psi-P)^2\right],
\eea
which is the form given in the introduction.

\section{The $\mathcal{N}=1$ M-theory solution}
Again, we begin with a review of the $AdS$ geometry. The MN(II) metric is
\bea\label{mn1}
\dd s^2&=&\frac{1}{\l}\left[\dd s^2(AdS_5)+\frac{1}{3}\dd
  s^2(H^2)+\frac{1}{9}(1-\l^3\r^2)\Big(\dd
  s^2(S^2)+(\dd\psi+P-P')^2\Big)+\frac{\l^3}{4(1-\l^3\r^2)}\dd\rho^2\right],\nn&&
\eea
where now
\bea
\l&=&\frac{4}{4+\rho^2},\nn
\dd P&=&\mbox{Vol}[S^2],\nn
\dd P'&=&\mbox{Vol}[H^2].
\eea
This time, the range of $\rho$ is $[-2/\sqrt{3},2/\sqrt{3}]$; at
$\rho=\pm2/\sqrt{3}$, an $S^3$ smoothly degenerates. This manifold
admits eight Killing spinors, which collectively define an $SU(2)$
structure. If we define the frame
\bea
e^1+ie^2&=&\frac{1}{\sqrt{3\l}}e^{i\gamma\psi}(\dd\mu+i\sinh\mu\dd\beta),\nn
e^3+ie^4&=&\frac{1}{3}\sqrt{\frac{1-\l^3\r^2}{\l}}e^{i\delta\psi}(\dd\theta+i\sin\theta\dd\phi),\nn
e^5&=&\frac{1}{3}\sqrt{\frac{1-\l^3\r^2}{\l}}(\dd\psi+P-P'),\nn
\hat{\rho}&=&\frac{\l}{2\sqrt{1-\l^3\r^2}}\dd\r,
\eea
where the constant phases $\gamma,\delta$ sum to unity,  then the $SU(2)$
structure forms are given by
\bea
J_4&=&e^{12}+e^{34},\nn
\Omega_4&=&(e^1+ie^2)(e^3+ie^4).
\eea
It may be explicitly verified that this six-dimensional $SU(2)$
structure satisfies the conditions of \cite{ypq}.

The MN(II) solution is interpreted as coming from M5 branes wrapping a
$H^2$ K\"{a}hler two-cycle in a Calabi-Yau three-fold. Again, we will
make this more precise, by exhibiting MN(II) as a solution of the
1/8 BPS $SU(3)$ analogue of the Fayazzuddin-Smith equations. In this
case, half the Killing spinors of the $AdS$ manifold define an $SU(3)$
structure, with 
structure forms $J_6$, $\Omega_6$. Then the supergravity description of
$1/8$ BPS M5 branes wrapping a K\"{a}hler two-cycle in a Calabi-Yau
three-fold \cite{brinne}, \cite{wrap} is as follows. The metric and flux are
\bea
\dd s^2&=&L^{-1}\dd s^2(\mbb^{1,3})+\dd s^2(\mathcal{M}_6)+L^2\dd t^2,\nn
\star_7F&=&-L^2\dd(L^{-2}J_6),
\eea
where $\mathcal{M}_6$ admits a globally-defined $SU(3)$ structure, and
again, all conventions and orientations follow \cite{wrap}. The
torsion conditions for the structure are
\bea
\dd t\w\dd(L^{-1}J\w J)&=&0,\nn
\dd(L^{-3/2}\Omega)&=&0.
\eea
These, together with the Bianchi identity, are sufficient to guarantee
a solution of eleven-dimensional supergravity. 

We now perform the frame rotation exactly as in the previous
section. The relationship between the Minkowski and $AdS$ frames is
\bea
e^a_{Mink}&=&e^a_{AdS},\nn
e^6&=&\cos\theta\hat{\rho}+\sin\theta\hat{r},\nn
\hat{t}&=&-\sin\theta\hat{\r}+\cos\theta\hat{r},
\eea
where now $a=1,...,5$. Again, $\l^{3/2}\rho=\cos\theta$. Therefore, $\hat{r}$ is
anti-aligned with $\hat{t}$ at $\rho=-2/\sqrt{3}$. It then rotates through
an angle of $\pi$ as $\rho$ spans its range, so that it is aligned
with $\hat{t}$ at $\rho=2/\sqrt{3}$. We find that $e^6,
\hat{t}$ are given by
\bea
e^6&=&Le^{-r/2}\dd\left(-\frac{1}{3}e^{-3r/2}\sqrt{4-3\r^2}\right),\nn
\hat{t}&=&L\dd\left(-\frac{\rho}{2}e^{-2r}\right).
\eea
Defining the Minkowski-frame coordinates,
\bea
u&=&-\frac{1}{3}e^{-3r/2}\sqrt{4-3\r^2},\nn
t&=&-\frac{\rho}{2}e^{-2r},
\eea
the metric in the Minkowski frame is given by
\bea
\dd s^2&=&L^{-1}\left[\dd s^2(\mbb^{1,3})+\frac{F^2}{3}\dd
s^2(H^2)\right]+L^2\left[F^{-1}\Big(\dd u^2+\frac{u^2}{4}[\dd
s^2(S^2)+(\dd\psi+P-P')^2]\Big)+\dd t^2\right],\nn&&
\eea
where $F=e^r$. This time, in order to determine $F$ in terms of the
Minkowski-frame coordinates, we must find the roots of a {\it quartic}
polynomial. The polynomial is
\bea\label{quartice}
12t^2e^{4r}+9u^2e^{3r}-4=0.
\eea
The Minkowski frame is given by
\bea
e^1+ie^2&=&\frac{F}{\sqrt{3L}}e^{i\gamma\psi}(\dd\mu+i\sinh\mu\dd\beta),\nn
e^3+ie^4&=&-\frac{Lu}{2\sqrt{F}}e^{i\delta\psi}(\dd\theta+i\sin\theta\dd\phi),\nn
e^5&=&-\frac{Lu}{2\sqrt{F}}(\dd\psi+P-P'),\nn
e^6&=&\frac{L}{\sqrt{F}}\dd u,\nn
\hat{t}&=&Ldt.
\eea
Again, the minus signs come from the definition of $u$. Then the
$SU(3)$ structure of MN(II) is given by
\bea
J_6&=&e^{12}+e^{34}+e^{56},\nn
\Omega_6&=&(e^1+ie^2)(e^3+ie^4)(e^5+ie^6).
\eea
At this point, repeating the analysis of the previous section
directly, we would let $L,F$ become arbitrary functions of $u,t$, and
then find the general Calabi-Yau solution. The torsion conditions and
Bianchi identity reduce to
\bea
\partial_t^2F+\frac{1}{u}\partial_u(uF\partial_uF)&=&0,\nn
L^3+\frac{2F^2}{3u}\partial_uF&=&0.
\eea
It follows from the construction of
\cite{ypq}, \cite{wrap} that the root of the quartic corresponding to the
MN(II) solution solves these equations. It seems very likely that so
do all the roots, though this has not been verified. However, it turns
out that there is no Calabi-Yau solution. To find one, we must extend
the ansatz, to  
\bea
\dd s^2&=&L^{-1}\left[\dd s^2(\mbb^{1,3})+\frac{F_1F_2}{3}\dd
  s^2(H^2)\right]\nn&&+L^2\left[F_1^{-1}\left(\dd
    u^2+\frac{u^2}{4}(\dd\psi+P-P')^2\right)+ F_2^{-1}\frac{u^2}{4}\dd
  s^2(S^2)+\dd t^2\right].
\eea
This extension of the ansatz is not unnatural as it clearly contains
MN(II) as the special case $F_1=F_2$. Furthermore it leaves the
$(3,0)$ form $\Omega$ invariant\footnote{Observe that $L^{-3/2}\Omega$ is always closed
with this frame ansatz.}; it is a purely K\"{a}hler deformation
of the $SU(3)$ structure. We also make the obvious modification of the
frame ansatz. In general, the torsion conditions and Bianchi identity
are rather complicated. However it is easy to determine the most
general Calabi-Yau solution with this ansatz, imposing closure of $J$
and constancy of $L$. The Calabi-Yau condition reads
\bea
\partial_tF_1=\partial_tF_2&=&0,\nn
 \frac{1}{3}\partial_u(F_1F_2)+\frac{u}{2F_1}&=&0,\nn
\partial_u\left(\frac{u^2}{4F_2}\right)-\frac{u}{2F_1}&=&0,
\eea
with general (positive signature) solution
\bea
F_1&=&\frac{3a^4}{u^2}\cos^2\xi,\nn
F_2&=&\frac{u^2}{2a^2}\frac{(1+\sin\xi)}{\cos^2\xi},
\eea
where $a^2,b$ are constants and $\sin\xi$ is a root of the cubic equation
\bea
-\frac{1}{3}\sin^3\xi+\sin\xi=b-\frac{u^4}{12a^6}.
\eea
This Calabi-Yau has two moduli. One, as usual, is just the overall
scale. Defining 
\bea
R=\frac{u^2}{2\sqrt{3}a^3},
\eea 
the metric is
\bea
\dd s^2&=&a^2\left[\frac{1}{2}(1+\sin\xi)\dd
s^2(H^2)+\frac{\cos^2\xi}{2(1+\sin\xi)}\dd
s^2(S^2)+\frac{1}{\cos^2\xi}\Big(\dd
R^2+R^2(\dd\psi+P-P')^2\Big)\right], \nonumber
\eea
\bea\label{218}
-\frac{1}{3}\sin^3\xi+\sin\xi=b-R^2.
\eea
The modulus $b$ parameterises inequivalent metrics. The generic metric
of this form has three degeneration points: $R=0$ and
$\xi=\pm\pi/2$. The point $\xi=-\pi/2$ (where the $H^2$ cycle
degenerates) is necessarily and expectedly singular. This is where
fractional branes are wrapped, in the probe picture. If $R=0$ and
$\xi=\pi/2$ do not coincide, the
point $R=0$ is also singular, since there the $U(1)$ degenerates with
a $4\pi$ periodicity inherited from the $AdS$ frame. To analyse what
happens near $\xi=\pi/2$, we expand the cubic to fourth order in $\xi$
in the vicinity of this point to find
\bea
R^2=\left(b-\frac{2}{3}\right)+\frac{\xi^4}{4}.
\eea
Clearly we require $b\ge 2/3$ (otherwise $\xi=\pi/2$ is not part of
the space). If $b>2/3$, then $R^2$ goes to a fixed positive value at
$\xi=\pi/2$; the metric there is clearly singular when written in
terms of $\xi$. However the point in moduli space $b=2/3$ where the
$R=0$ and $\xi=\pi/2$ degeneration points of the metric coincide is
special. With $b=2/3$ the metric near $\xi=\pi/2$ becomes
\bea
\dd s^2=a^2\left[\dd s^2(H^2)+\dd\xi^2+\frac{\xi^2}{4}\Big(\dd s^2(S^2)+(\dd\psi+P-P')^2\Big)\right].
\eea
With the periodicity of $\psi$ as inherited from the $AdS$ frame, an
$S^3$ smoothly degenerates. The Calabi-Yau \eqref{218}, with $b=2/3$,
is interpreted as the image of MN(II) under an inverse geometric transition. Analysing the relationship between $R$
and $\xi$ near $\xi=-\pi/2$, we can deduce that the range of $R$ is
is either $[-2/\sqrt{3},0)$ or $(0,2/\sqrt{3}]$. In either branch, the
singularity is at finite proper distance.

\section{Conclusions and outlook}
In this paper a way of mapping a supersymmetric $AdS$ manifold to a
special holonomy manifold has been proposed. The main
conclusion is that this procedure should be applicable
to all known wrapped-brane $AdS$ solutions of string and M-theory; it
will be very interesting to explore the the special holonomy metrics
in each case. Given the metrics for the string theory solutions, it
should be possible to make progress towards constructing the dual
field theories. 

The construction relies in an essential way on the existence of an
interpolating solution. For MN(I) and (II), we can say a little about what
the boundary conditions for an interpolation should be. For MN(I), the interpolating
solution should be globally smooth, and should contain a neighbourhood
where the metric is 
diffeomorphic to the limit of the metric \eqref{eguchi} as
$R\rightarrow1$. It should also contain a neighbourhood where the
metric is diffeomorphic to the limit as $\rho\rightarrow0$ of
\eqref{mn1}. For MN(II), an interpolating solution should contain a
neighbourhood diffeomorphic to the Calabi-Yau metric \eqref{218} near
$\xi=\pi/2$. It should also contain a neighbourhood where the metric
is diffeomorphic to \eqref{mn1} as
$\rho\rightarrow\pm2/\sqrt{3}$. Global topological considerations will
be important in trying to construct an interpolation; for example, by
a careful analysis it should be possible to fix the relative scales of
the MN/CY metrics. This might be done, for example, by comparing the
sizes of the $H^2$ bolts in the $AdS$ and Calabi-Yau metrics, at the
point in each where the $U(1)$ or $S^3$ degenerates, for MN(I) and MN(II)
respectively. However since an interpolating solution would
necessarily be cohomogeneity two, and the governing equations are
non-linear, finding one explicitly will be challenging. It might be
worthwile to perform a numerical analysis, if a better handle can be
obtained on the boundary conditions.

There might be other, more complicated, Calabi-Yau manifolds that
could be related to the Maldacena-Nu\~{n}ez solutions. This would be
analagous to the way in which conical Calabi-Yaus can be
thought of as generic local models for a particular sort of
singularity, in a manifold whose global structure could be much more
complicated. Placing D-branes at the singularity is usually argued to
produce an $AdS$ throat, which is insensitive to the global
structure. It would be interesting to know if the Calabi-Yau metrics
obtained here can be thought of in a similar way - as local models of
a more generic type of singularity. The topology of the manifolds in
this paper is more complicated, so it is not obvious yet whether or
not this is true. In any event, for the
purposes of constructing the dual, in the conical case only the
geometry near the singularity - the conical metric - is
required. Analagously, for the purposes of constructing the duals of
the Maldacena-Nu\~{n}ez solutions, the metrics of this paper are
interpreted as the appropriate backgrounds.

There appears to be an intriguing link between solutions of the
various non-linear equations we have encountered and roots of
polynomials. This seems to suggest 
some underlying algebraic geometry which has not been properly
appreciated. It will be interesting to explore this in more detail; it
appears to be a generic feature of how $AdS$ manifolds solve wrapped
brane structure equations.  

\section{Acknowledgements}
I would like to express my gratitude to Pau Figueras and Jerome
Gauntlett for collaboration during different stages of this
project. This work was supported by EPSRC.

\end{document}